\documentclass[12pt]{trans2-l}
\usepackage{amsmath, amssymb,amsthm, latexsym, times, graphicx,citesort}

\newtheorem{theorem}{Theorem}[section]
\newtheorem{lemma}[theorem]{Lemma}

\theoremstyle{definition}
\newtheorem{definition}[theorem]{Definition}

\theoremstyle{remark}
\newtheorem{remark}[theorem]{Remark}

\numberwithin{equation}{section}
\def\C{\Bbb{C}}
\def\R{\Bbb{R}}

\def\CC{\Bbb{C}}
\def\RR{\Bbb{R}}

\def\mS{\mathcal{S}}

\newcommand{\norm}[1]{\ensuremath{\parallel\!\! {#1}\!\!\parallel} }
\newcommand{\dpr}{\ensuremath{\prime\prime}}

\begin{document}
\title{On guided electromagnetic waves in photonic crystal waveguides}

\author{Peter Kuchment}
\address{Department of Mathematics, Texas A\&M University, College Station, TX 77843-3368}
\email{kuchment@math.tamu.edu}
\author{Beng-Seong Ong}
\address{Veritas, Houston, TX}
\email{beng.ong@cggveritas.com}
\thanks{Work of both authors was partially supported by the NSF Grants DMS 0296150 and 9971674. Work of the first author was also supported by the NSF Grant DMS 0072248.}


\dedicatory{Dedicated to the memory of Professor V. B. Lidskii}



\begin{abstract}
The paper addresses the issue of existence and confinement of electromagnetic modes guided by linear defects in photonic crystals. Sufficient condition are provided for existence of such waves near a given spectral location. Confinement to the guide is achieved due to a photonic band gap in the bulk dielectric medium.
\end{abstract}

\maketitle

\section{Introduction}

A photonic crystal, also called photonic band-gap (PBG) material, is a periodic medium which plays the role of an optical analog of a semi-conductor. Such a medium has a gap in the frequency spectrum of electromagnetic (EM) waves. The idea of a photonic crystal was first suggested in 1987
\cite{J87,Y}, and has since been intensively studied experimentally and theoretically (see, e.g.,
the recent books \cite{JMW,JJ,Sakoda,Skorob,Zolla}, the mathematical survey \cite{Ku_pbg}, the on-line bibliography \cite{bibl}, and references therein). This interest has been triggered by the numerous promising applications of PBG materials, one of which is using photonic crystal for manufacturing highly efficient optical waveguides. The idea is to introduce a linear ``defect'' into a PBG material, and to guide through it EM waves of a frequency prohibited in the bulk. Numerical and experimental studies have shown that such superior guides can be efficiently created, e.g. \cite{JMW,JJ,LCM,Sakoda,Skorob}.

In order to create such a guide, one needs to establish several facts. The first, and foremost, is existence of guided waves of frequencies in the band gap. The second, and an easier one, is confinement of these modes. This paper addresses both issues, by finding some sufficient conditions of existence of guided modes, and showing their confinement to the guide, in the sense of being evanescent in the balk. Similar results were previously obtained by the authors in \cite{Ku_Ong} for scalar models (i.e., for acoustic analogs of PBG waveguides). In this paper, we will address the above questions for the full Maxwell case.

There is another important question to be resolved. Namely, one needs to show that the impurity spectrum that arises in the spectral gaps due to the presence of a linear defect does not correspond to bound states. This difficult issue is not addressed here (see some relevant remarks and references in Section \ref{S:remarks}).

In Section \ref{S:prelim} we introduce the main model to be investigated. The next Section \ref{S:results} contains formulation of the main results, with the proofs provided in Section \ref{S:proofs}. The paper ends with the sections devoted to final Remarks and Acknowledgments, as well as the Bibliography.

\section{Preliminaries}\label{S:prelim}

We start by describing the mathematical model studied in this paper. Let $\varepsilon_0 (x)$ be a bounded positive measurable functions in $\R^3$ separated from zero:

\begin{equation}\label{E:e0}
0<c_0 \leq \varepsilon_0(x)
\leq c_1 < \infty.
\end{equation}

It is usually assumed in photonic crystal
theory that $\varepsilon_0$ is periodic with respect to a lattice $\Gamma \subset \R^3$, but this is not required for our results.

The function $\varepsilon_0$ represents the dielectric properties of the bulk material. In other words, one can think of the space $\R^3$ filled with a dielectric material with the dielectric function $\varepsilon_0$.

The unperturbed Maxwell operator $M_0$ is the self-adjoint
realization of
\begin{equation}
M_0:=\nabla^\times \frac{1}{\varepsilon_0(x)} \nabla^\times
\end{equation}
in $L_2(\Bbb{R}^3;\Bbb{C}^3)$ defined by means of its quadratic
form
\begin{equation} \int
\varepsilon_0^{-1}|\nabla^\times u |^2 dx
\end{equation}
with the domain $H^1(\Bbb{R}^3;\Bbb{C}^3)$.
We use here the shorthand notation
$$
\nabla^\times u = \nabla \times u =
\mbox{curl}\ u.
$$

\begin{figure}[ht!]\begin{center}
\scalebox{.5}{\includegraphics{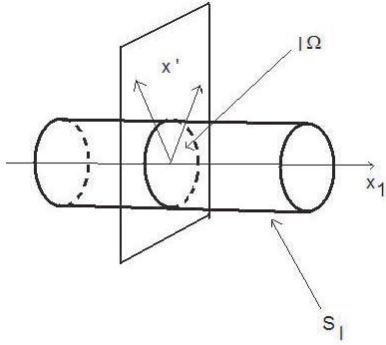}}
\caption{The waveguide $\mathcal{S}_l$.}\label{F:strip}
\end{center}
\end{figure}

A cylindrical domain $\mathcal{S}_l$ (see Fig. \ref{F:strip}) will represent a linear  ``defect strip'':
$$
\mathcal{S}_l:=\{x=(x_1,x^\prime) \in \Bbb{R}^3 \, | \, x_1 \in
\Bbb{R}, \, x^\prime \in l\Omega\}.
$$
Here the cross-section $l\Omega$ of the strip is a domain $\Omega$ in $\Bbb{R}^{2}$ (e.g., the unit ball
centered at the origin), scaled with factor $l$.

We now introduce the perturbed medium with homogeneous dielectric properties inside the defect strip $\mS_l$:
\begin{equation}
\varepsilon(x)=\left\{
\begin{array}{cc}
  \varepsilon > 0 & \mbox{for } x \in \mathcal{S}_l  \\
  \varepsilon_0(x) & \mbox{for } x \notin \mathcal{S}_l \\
\end{array}\right..
\end{equation}

The perturbed Maxwell operator
\begin{equation}
M:=\nabla^\times \frac{1}{\varepsilon(x)} \nabla^\times
\end{equation}
corresponds to the medium with the linear defect. It is defined, analogously to $M_0$, as a self-adjoint operator in $L_2(\RR^3;\CC^3)$.

\begin{remark}
The reader has probably noticed that we disregard the standard restriction of the Maxwell operator to divergence-free fields \cite{JMW}. The difference is essentially in acquiring a huge eigenspace corresponding to the zero frequency, with all other parts of the spectral decomposition staying intact. Since the problem of guided waves concerns the situation inside the spectral gaps of the operator $M_0$, this difference is irrelevant in this case. On the other hand, abandoning the zero divergence condition will simplify the techniques considerably.
\end{remark}

Our goal is the same as in the paper \cite{Ku_Ong} devoted to the scalar (acoustic) case, i.e. to show
that for any gap $(\alpha, \beta)$ in the spectrum $\sigma(M_0)$ of the unperturbed medium, under appropriate conditions on the parameters $l$ and $\varepsilon$ of the line defect, additional spectrum arises inside the gap, with the corresponding (generalized) eigenmodes being confined to the defect (evanescent in the bulk).

\section{Formulation of the results}\label{S:results}

In order to formulate our first main result, we need to introduce the following quantity:
\begin{definition}
We denote by $\nu>0$ the lowest eigenvalue of the Laplace operator $\Delta$ acting on divergence free $\R^2$-valued vector fields on $\Omega$ with Dirichlet boundary conditions on $\partial \Omega$.

I.e.,  $\nu>0$ is the smallest number for each a non-trivial solution of the following problem exists:
\begin{equation}\label{E:nu}
    \begin{cases}
    \Delta E=\nu E \mbox{ in } \Omega\\
    \nabla\cdot E=0 \mbox{ in } \Omega\\
    E\mid_{\partial \Omega}=0,\\
    \end{cases}
\end{equation}
where $\frac{\partial }{\partial n}$ is the external normal derivative on $\partial \Omega$.
\end{definition}

Our main results are given in the following theorems.

\begin{theorem}\label{TM:existence}
Let $G=(\alpha,\beta)$ be a non-empty finite interval, such that
$\alpha>0$ (we will be especially interested in the case when $G$ is a gap in the spectrum of the ``background medium'' operator $M_0$). Let the following inequality be satisfied:
\begin{equation}
l^2 (\beta-\alpha) \varepsilon > 2\nu . \label{EM:condition}
\end{equation}
Then the interval $G$ contains at least one point of the spectrum $\sigma(M)$ of the perturbed operator.
\end{theorem}
This theorem guarantees that when (\ref{EM:condition}) is
satisfied, eigenmodes of the perturbed medium do arise in the spectral gaps of the background medium. Furthermore, if $\delta>0$ is such $l^2 \delta \varepsilon > \nu$, the corresponding spectrum forms a $\delta$-net in the gap.

Before one can fully associate these modes with the guided waves, one needs to establish their confinement to the waveguide (i.e., their evanescent nature in the bulk of the material). In order to describe the corresponding result, we need to remind the reader some notions and results about generalized eigenfunction expansions. Here by generalized eigenfunctions one understands solutions of the eigenvalue problem, which do not decay sufficiently fast (or do not decay at all) to belong to the ambient Hilbert space $L^2$. One can find detailed discussions of generalized eigenfunction expansions, for instance, in \cite{Ber,Shubin}.

Namely, as it is shown in \cite{Klein}, for the operator $M$ that we consider, for almost any (with respect to the spectral measure) $\lambda$ in the spectrum $\sigma(M)$, there is a generalized eigenfunction $u_\lambda(x)\in H^1_{loc}(\R^3,\C^3)$ with the growth estimates
\begin{equation}
(1+|x|)^{-N} u(x) \in L_2(\RR^3;\CC^3) \;\;\mbox{and}\;\;
(1+|x|)^{-N}\, \nabla^\times u(x) \in L_2(\RR^3;\CC^3)
\label{EM:polybound}
\end{equation}
for some $N>0$. This system of generalized functions is complete in the whole space. For elliptic operators with smooth coefficients, this is a well known fact \cite{Ber}.

\begin{definition}
We refer to the following growth condition as \emph{polynomial boundedness of order
$N$}: for any compact set $K\subset \RR^3$ and $x\in \RR^3$,
\begin{equation}
\|u\|_{L_2((K+x);\CC^3)}+\|\nabla^{\times}
u\|_{L_2((K+x);\CC^3)}\leq C_K(1+|x|)^N \label{EM:growth}
\end{equation}
\end{definition}

In the next result we will use the following notation: for $x =(x_1,
x^{\prime})\in \RR^3$, we denote by $\chi_x(y)$ the {\emph characteristic
function} of the cube 
$$
\{y\mid |y_j-x_j|\leq 1 \mbox{ for }j=1,2,3\}
$$ 
centered at $x$. I.e., $\chi_x(y)$ is equal to $1$ when $y$ is in this cube, and $0$ otherwise.

\begin{theorem}\label{TM:decay}
Let $G$ be a finite spectral gap of $M_0$ and $u_\lambda$ be a polynomially bounded generalized eigenfunction of $M$
corresponding to $\lambda \in G \cap \sigma(M)$. Then there exist
positive constants $C$ and $C(\lambda)$ such that
\begin{equation}
\|\chi_x u_\lambda\|\leq C \left(1+|x_1|\right)^N e^{-C(\lambda)
dist(x,\,S_l)}, \label{EM:decay}
\end{equation}
where $N$ is the order of polynomial boundedness of $u_\lambda$.
\end{theorem}

When the bulk medium is periodic in the $x_1$-direction, the polynomial growth in (\ref{EM:decay})
disappears:

\begin{theorem}\label{TM:periodic_decay}
If $\varepsilon_0(x)$ is periodic in the $x_1$-direction, then one can find a complete family of generalized eigenfunctions that satisfies
\begin{equation}
\|u_\lambda\|_{L_2((K+x);\CC^3)}+\|\nabla^{\times}
u_\lambda\|_{L_2((K+x);\CC^3)}\leq C_K(1+|x^\prime|)^N
\label{EM:periodic_growth}
\end{equation}
for any compact set $K\subset \RR^3$, $x\in \RR^3$. In this case, for $\lambda\in G\cap \sigma(M)$, one has the estimate
\begin{equation}
\norm{\chi_x u_\lambda} \leq C e^{-C(\lambda)\mbox{dist}(x, \mathcal{S}_l)}.
\label{EM:periodic_decay}
\end{equation}
\end{theorem}

\section{Proofs of the results}\label{S:proofs}

In what follows, the norm and inner product in
$L_2(\Bbb{R}^n;\Bbb{C}^3)$ will be denoted by $\|\cdot\|$ and $\langle\cdot,\cdot\rangle$ respectively.

\subsection{Proof of Theorem \ref{TM:existence}}

We will show that if $\mu>0$ and $\delta>0$ is such that $l^2\delta\epsilon>\nu$, there is spectrum of the operator $M$ in the $\delta$-vicinity of $\mu$. Then, taking $\mu=(\alpha+\beta)/2$ and $\delta=(\beta-\alpha)/2$, one gets the statement of the theorem.

In order to show this, due to self-adjointness of $M$, it is sufficient to find a vector function $w\in L_2(\RR^3,\CC^3)$ of unit norm, such that
\begin{equation}
\norm{M w-\mu w}^2<\delta^2. \label{EM:approx}
\end{equation}

Let $g$ be a smooth, unit $L^2$-norm, divergence free real vector field on $\Bbb{R}^2$ with compact support in $\Omega$ and unit
$L^2$-norm, i.e, $g(y,z) = (\phi(y,z),\zeta(y,z))$ where $\phi$,
$\zeta \in C_0^\infty(\Omega)$ and $\nabla\cdot g =0$. We define
$$
g_l:=(\phi_l,\zeta_l)=l^{-1}(\phi(x^\prime/l),\zeta(x^\prime/l)).
$$
Then $g_l$ is also a unit $L^2$-norm divergence free field. Let $\psi(x_1) \in C_0^\infty(\Bbb{R})$ have unit $L_2(\R)$-norm and
$\psi_n(x_1)=n^{-1/2}\psi(x_1/n)$ for $n>0$. Clearly
$\psi_n(x_1)$ also has unit $L_2$-norm.

Denoting $k=\sqrt{\mu\varepsilon}$, we introduce the following candidate for an approximate eigenfunction:
\begin{equation}
w_{l,n}(x)=\psi_n(x_1)e^{ikx_1} \left(
\begin{array}{c}0\\ \phi_l(x^\prime)\\ \zeta_l(x^\prime) \end{array} \right).
\end{equation}
The function $w_{l,n}$ clearly has unit norm in $L_2(\RR\times
l\Omega)$.

Instead of estimating the left hand side of (\ref{EM:approx}), we
will estimate \\ $\norm{\varepsilon\left(M w-\mu w \right)}^2$.
Taking into account that the function $w$ is supported inside the
defect, the needed inequality (\ref{EM:approx}) can be also
rewritten as
\begin{equation}
\norm{\nabla^\times \nabla^\times w- k^2 w}^2<\delta^2
\varepsilon^2. \label{EM:approx2}
\end{equation}

Using the identity
$$\nabla^\times \nabla^\times w= -\Delta w +
\nabla (\nabla \cdot w )$$ and that $g_l$ is divergence free, we
obtain
$$\norm{\nabla^\times \nabla^\times w-k^2 w}^2 =
{\left|\left|
\begin{pmatrix}
0 \\
-(\psi_n^{\dpr}+2ik\psi_n^\prime)\phi_l - \psi_n \Delta \phi_l\\
-(\psi_n^{\dpr}+2ik\psi_n^\prime)\zeta_l - \psi_n \Delta \zeta_l\\
\end{pmatrix}  \right|\right|}^2
$$
where the norms are in $L_2(\RR^3;\CC^3)$.

Since the functions $\phi$, $\zeta$, and $\psi$ are real
valued, their assumed normalization shows that the above expression is equal to
$$
\begin{array}{c}
n^{-4}\norm{\psi^{\dpr}}_{L_2(\Bbb{R})}^2
+4k^2n^{-2}\norm{\psi^\prime}_{L_2(\Bbb{R})}^2
+l^{-4}\norm{\Delta g}_{L_2(\Omega,\R^2)}^2\\
+2(nl)^{-2}\langle\psi^{\prime\prime},\psi
\rangle_{L_2(\Bbb{R})}\langle\Delta g ,g\rangle_{L_2(\Omega,\R^2)}.
\end{array}
$$
Since $n$ can be chosen arbitrarily large, the terms with the factors that are negative powers of $n$ can be made
arbitrarily small (uniformly with respect to $k$ on any finite interval). Hence, one needs to control only the remaining terms by an appropriate choice of a divergence free vector field $g$. In other words, one is interested in making $l^{-4}\norm{\Delta g}_{L_2(\Omega,\R^2)}^2$ smaller than $\delta^2 \varepsilon^2$, i.e.
\begin{equation}
\norm{\Delta g}_{L_2(\Omega;\Bbb{R}^2)}<l^2 \delta
\varepsilon, \label{EM:approx4}
\end{equation}
while keeping $\langle\Delta g, g\rangle$ under control.

Since
$$
\nu = \inf \norm{\Delta g}_{L_2(\Omega;\Bbb{R}^2)},
$$
where the \textit{infimum} is taken over real, unit $L^2$-norm, divergence free vector fields  $g\in C_0^\infty(\Omega;\RR^2)$, this condition boils down to
\begin{equation}
l^2\delta \varepsilon > \nu,
\end{equation}
which proves the statement of the theorem.\qed

\subsection{Proof of Theorem \ref{TM:decay}}

We assume here that $\lambda$ belongs to a finite gap $G$ of the spectrum of $M_0$ and $u:=u_\lambda$ is the corresponding generalized eigenfunction of $M$ satisfying (\ref{EM:growth}). Let us also introduce the resolvent $R(\lambda)=(M_0-\lambda)^{-1}$. We will also use the function $\chi_x$ introduced before Theorem \ref{TM:decay}.

We will need the following auxiliary statement concerning the exponential decay of the resolvent, which is a result of \cite{FKl_EM}:

\begin{lemma}\cite{FKl_EM}\label{L:resolvent_est}
There exists a positive number $m_\lambda$ that depends only on the
distance of the point $\lambda$ from the gap edges, such that for a positive constant $C$, the following estimates hold for the local
$L_2 (\Bbb{R}^3;\Bbb{C}^3)$-norm of the resolvent $R(\lambda)$:
\begin{equation}
\begin{array}{c}
\norm{\chi_u R(\lambda) \chi_v} \leq C e^{-m_\lambda |u-v|} \\
\norm{\chi_u \nabla^\times R(\lambda) \chi_v} \leq C e^{-m_\lambda |u-v|} \\
\end{array}
\label{E:resolvent}
\end{equation}
for any $u,\,v \in \Bbb{R}^3$. Here the norms in the left hand
side are the operator norms in $L_2(\Bbb{R}^3;\Bbb{C}^3)$.
\end{lemma}

We now consider the sesqui-linear form
$$ Q[\varphi,w]:= \langle \nabla^\times \varphi,
\frac{1}{\varepsilon_0}\nabla^\times w \rangle -\lambda \langle\varphi,
w\rangle
$$ with the domain $H^1(\Bbb{R}^3;\Bbb{C}^3)$.

Let $\varphi := R(\lambda)\chi_x u$. Note that $\varphi$ belongs to the domain of the operator $M_0$.

Let $p = \max\left(2\mbox{dist}(x,S_l),1\right)$ and $\xi_{x}(y)$
be a nonnegative smooth cutoff function that depends on $y_1$
only, is supported in $(x_1-(p+1),x_1+(p+1))$ and such that it is
equal to $1$ on $[x_1-p,x_1+p]$. We assume further that
$\xi_{x}(y)\leq 1$ and $|\nabla\xi_{x}(y)|\leq C$ for some constant $C$ and all $x,y\in \Bbb{R}^3$.
Note that $\xi_x u \in H^1(\Bbb{R}^3;\Bbb{C}^3)$.
Using $w=\xi_x u$, one gets
$$
Q[\varphi,\xi_x u]= \langle M_0\varphi, \xi_x u \rangle- \langle
\lambda\varphi, \xi_x u \rangle = \langle \chi_x u, \xi_x u \rangle =
\norm{\chi_x u}^2.
$$
Thus, the goal is to estimate $Q[\varphi,\xi_x u]$
from above. On the other hand, using the equality $Mu=\lambda u$ and
integration by parts, one obtains
\begin{equation}\label{E:form}
\begin{array}{c}
  Q[\varphi,\xi_x u]
=\langle \nabla^\times \varphi,
\varepsilon_0^{-1}\nabla^\times (\xi_x u)\rangle-
\langle\varphi, \xi_x \lambda u \rangle\\
  = \langle\nabla^\times \varphi, \tilde{\epsilon}\nabla^\times (\xi_x
  u)\rangle + \langle\nabla^\times \varphi, \varepsilon^{-1}\nabla^\times (\xi_x
  u)\rangle  - \langle\varphi, \xi_x Mu\rangle   \\
= \langle\nabla^\times \varphi, \tilde{\epsilon}\nabla^\times (\xi_x
  u)\rangle + \langle\nabla^\times \varphi, \varepsilon^{-1}\nabla^\times (\xi_x
  u)\rangle  - \langle \nabla^\times(\xi_x \varphi), \varepsilon^{-1}\nabla^\times u\rangle,
\end{array}
\end{equation}
where we used the notation
$$
\tilde{\epsilon}(x):=\frac{1}{\varepsilon_0(x)}-\frac{1}{\varepsilon(x)}.
$$
Notice that $\tilde{\epsilon}$ is supported inside the strip $\mathcal{S}_l$.

Using the identity $\nabla^\times (\xi
u)=\xi\nabla^\times u + \nabla \xi \times u$, the first two terms
in the last line of (\ref{E:form}) can be combined to obtain
$$\begin{array}{c}
\langle\nabla^\times \varphi, \tilde{\epsilon}(\xi_x \nabla^\times u
+ \nabla \xi_x \times u) \rangle + \langle\nabla^\times \varphi,
\varepsilon^{-1}(\xi_x \nabla^\times u + \nabla \xi_x \times u)
\rangle\\
=\langle\nabla^\times \varphi, \xi_x \tilde{\epsilon}\nabla^\times
u\rangle + \langle\nabla^\times \varphi,
(\tilde{\epsilon}+\varepsilon^{-1}) \nabla \xi_x \times u
\rangle +\langle\nabla^\times \varphi,
\varepsilon^{-1}\xi_x\nabla^\times u \rangle
\end{array}
$$
The term $- \langle \nabla^\times(\xi_x \varphi),
\varepsilon^{-1}\nabla^\times u\rangle$ can be expanded to
$$-\langle \xi_x\nabla^\times\varphi,\varepsilon^{-1}\nabla^\times
u\rangle-\langle \nabla \xi_x
\times\varphi,\varepsilon^{-1}\nabla^\times u\rangle
$$
Combining the last two expressions, we get
\begin{equation}
\langle\nabla^\times \varphi, \xi_x \tilde{\epsilon}\nabla^\times
u\rangle+\langle \nabla^\times
\varphi,{\varepsilon^{-1}_0}\nabla \xi_x \times u \rangle -
\langle \nabla \xi_x \times
\varphi,\varepsilon^{-1}\nabla^\times
u\rangle\label{EM:3terms}
\end{equation}

Our last task in proving the theorem is to estimate from above the
terms in (\ref{EM:3terms}). Let $V =
[x_1-p-1,x_1+p+1]\times l\Omega$. This is a compact domain that
can be covered by the union of $p$ fixed size domains $V_j =
[a_j,a_j+2]\times l\Omega$ and which contains the supports of
$(\xi \tilde{\epsilon})$. Also note that $\mbox{dist}(x,V_j)\geq
\mbox{dist}(x,S_l)$. Using Lemma \ref{L:resolvent_est} and (\ref{EM:growth}), we get for any $0<\eta<m_\lambda$
\begin{equation}
\begin{array}{c}
|\langle \nabla^\times \varphi, \xi_x \tilde{\epsilon}\nabla^\times
u\rangle| \leq \norm{\chi_{V}\nabla^\times \varphi}\,\norm{\xi_x
\tilde{\epsilon}\nabla^\times u}\\
\leq C \norm{\sum_j
\chi_{V_j}\nabla^\times R(\lambda)\chi_x u}\norm{\sum_j
\chi_{V_j}\nabla^\times u}\\
\leq C p^2(|x_1|+p+1)^{2N} e^{-m_\lambda \mbox{dist}(x,\mathcal{S}_l)}\\
\leq C(|x_1|+1)^{2N} e^{-(m_\lambda-\eta) \mbox{dist}(x,\mathcal{S}_l)}\\
\end{array}
\label{EM:est1}
\end{equation}
We used here that $p=\max(2\mbox{dist}(x,\mathcal{S}_l),1)$ and denoted by $C$ different constants.

Let us move now to estimating the last term in (\ref{EM:3terms}).
Denote by $a>0$ a number such that shifts of $l\Omega$ by vectors
$aj$ with $j\in \Bbb{Z}^{2}$ cover the whole space $\Bbb{R}^{2}$.
We denote
$$W_j :=
([x_1-p-1,x_1-p]\cup[x_1+p,x_1+p+1]) \times
\left(l\Omega+aj\right).$$ Then $W_j=W_0+(0,aj)$. Notice that
$W=\cup_j W_j$ covers $\mbox{supp}\, \nabla \xi$ and
$\mbox{dist}(x,W_j)\geq C_1(p+|j|)-C_2$.

We are now ready to estimate the last term of (\ref{EM:3terms})
from above. We proceed as before, using the lemma, the polynomial
growth of $u$, and uniform boundedness of $\nabla \xi_x$.
\begin{equation}
\begin{array}{c}
|\langle \nabla^\times \varphi,\varepsilon_0^{-1}\nabla
\xi_x^\times u \rangle|\leq C\sum\limits_j \norm{\chi_{W_j}u}
 \norm{\chi_{W_j}\nabla^\times R(z)\chi_x u} \\
\leq C \sum\limits_j (|x_1|+p+|j|+1)^{2N} e^{-m_\lambda \mbox{dist}(x,W_j)}\\
\leq C(|x_1|+p+1)^{2N} e^{-Cm_\lambda \mbox{dist}(x,\mathcal{S}_l)}\sum\limits_j%
(1+|j|)^{2N}e^{-Cm_\lambda |j|}\\
\leq C (|x_1|+1)^{2N} e^{-Cm_\lambda \mbox{dist}(x,\mathcal{S}_l)}. \\
\end{array}
\label{EM:est3}
\end{equation}

The middle term in (\ref{EM:3terms}) is estimated
analogously. Combining these estimates, we get
$$
\norm{\chi_x u}^2  = Q[\varphi,\xi_x
u] \leq C (1+|x_1|)^{2N}
e^{-Cm_\lambda\mbox{dist}(x,\mathcal{S}_l)}.
$$
This finishes the proof of the theorem.\qed

\subsection{Proof of Theorem \ref{TM:periodic_decay}}

In this periodic situation, operator $M$ has a complete family of generalized eigenfunctions that do not grow in the $x_1$-direction of periodicity. Indeed, according to Bloch-Floquet theory \cite{Ku_book,RS}, a generalized eigenfunction $u=u_\lambda$ of $M$ corresponding to $\lambda$ can be chosen as
$\tilde{u}(x)e^{i k x_1}$, where $\tilde{u}(x)$ is periodic in
$x_1$-direction with period $a$ and $k = 2\pi/a$. Thus, $u$ satisfies (\ref{EM:periodic_growth}). Then, repeating the previous proof, one comes up with the estimate (\ref{EM:periodic_decay}). \qed

\section{Remarks}\label{S:remarks}

\begin{enumerate}

\item Theorem \ref{TM:existence} provides sufficient conditions for the existence of a
$\delta$-net of the defect spectrum inside a spectral gap. One wonders
how much of the gap the guided mode spectrum can occupy. Can it fill the whole
gap ? If so, under what conditions ? There seem to be no rigorous results available concerning these questions.

\item Results of \cite{BCH} on improved Combes-Thomas resolvent
estimates show that the exponential decay constant $m_\lambda$, which
clearly depends on the distance of the point $\lambda$ from the
spectrum, behaves as $\sqrt{(\lambda-\alpha)(\beta-\lambda)}$ inside a gap $G=(\alpha,\beta)$.

\item As it has already being mentioned, in order to have the full
right to call the discovered modes ``guided'', one needs to show
that they do not correspond to point spectrum (i.e., to bound
states). Here the most treatable case should be of a periodic
medium with a linear defect aligned along one of the lattice
vectors. In this situation one can apply the Floquet-Bloch theory
with respect to the axial variable of the waveguide and hope to
use standard techniques applied in the case of Schr\"{o}dinger
operators with periodic potential (e.g., \cite{BSu,Fr_ac,Ku_book,RS}). This happens to be not an easy task. Even in the
case of ``hard wall'' periodic waveguides, when waves are
contained in a periodic waveguide by Dirichlet, Neumann, or more
general boundary conditions, this problem is non-trivial. Although
it has been considered for rather long time \cite{D,Ku_book}, the
first real advances are very recent \cite{Frank,Frank_Shter,Fr_guide,So,St,StSu}. The case of photonic crystall waveguides
is more complex, due to absence of complete confinement of the
waves, which exponentially decay into the bulk, but do not vanish
completely. Apparently, the most recent work devoted to this issue assumes the bulk material to be
homogeneous, with an embedded periodic guide \cite{Filonov}.
\end{enumerate}

\section{Acknowledgements}
The authors express their gratitude to H.~Ammari, D.~Dobson, P.~Exner, F.~Santosa, R.~Shterenberg, A.~Sobolev, and T.~Suslina for discussions and information.

This research was partly sponsored by the NSF through the Grants DMS 0296150, 0072248, and 9971674. The authors thank the NSF for the support.

\end{document}